\documentclass{aastex631}

\usepackage{amsthm,amsmath,amssymb}
\usepackage{lipsum}
\usepackage{float}
\usepackage{soul}

\begin{document}

\title{BSN: The First Multiband Light Curve Analysis of the W UMa-type Contact Binary System EM Tucanae}

\author[0009-0009-6128-8726]{Sepideh Houshiar}
\altaffiliation{sepideh.hoshyar@gmail.com}
\affiliation{Department of Physics, University of Qom, Qom, Iran}

\author[0000-0002-0196-9732]{Atila Poro}
\affiliation{LUX, Observatoire de Paris, CNRS, PSL, 61 Avenue de l'Observatoire, 75014 Paris, France}
\affiliation{BSN Project; Astronomy Department of the Raderon AI Lab., BC., Burnaby, Canada}

\author[0000-0001-9655-5796]{Abbas Abedini}
\affiliation{Department of Physics, University of Qom, Qom, Iran}

\author[0000-0002-9262-4456]{Eduardo Fernández Lajús}
\affiliation{Instituto de Astrofísica de La Plata (CCT La Plata-CONICET-UNLP), La Plata, Argentina}

\begin{abstract}
We present a comprehensive photometric light curve and orbital period analysis of the W UMa-type contact binary EM Tuc. The O-C analysis constructed from all available eclipse timings exhibits a clear upward parabolic trend, indicating a continuous increase in the orbital period at a rate of \(dP/dt = (1.401 \pm 0.042)\times10^{-7}\,\mathrm{d\,yr^{-1}}\). This behavior is consistent with mass transfer from the less massive to the more massive star in the system, and the corresponding mass-transfer rate is estimated to be \(\dot{M} = -(4.62 \pm 1.54)\times10^{-8}\,M_\odot\,\mathrm{yr^{-1}}\). Light curve modeling with the PHysics Of Eclipsing BinariEs (PHOEBE) Python code, refined through MCMC sampling, confirms an overcontact configuration and yields a mass ratio of $q = 3.987$. A cool photospheric starspot on the cooler component is required to reproduce the observed O'Connell asymmetry. Using Gaia DR3 parallax together with the photometric solution, the absolute masses of the components are derived as \(M_h = 0.24 \pm 0.05\,M_\odot\) and \(M_c = 0.97 \pm 0.19\,M_\odot\).
\end{abstract}

\keywords{binaries: eclipsing - methods: observational - stars: individual (EM Tuc)}

\section{Introduction}
W Ursae Majoris (W UMa) systems form a well-known subclass of eclipsing binaries in which both components fill their Roche lobes and exist in a state of geometric contact (\citealt{1971ARAA...9..183P}[1]). The two stars share a common convective envelope (\citealt{1968ApJ...151.1123L}[2]), allowing continuous transfer of mass and energy between them. As a result, although the components may differ in mass, structure, and evolutionary status, they maintain nearly identical surface temperatures, which is one of the defining observational characteristics of contact binaries. The orbital periods of W UMa systems are typically short, and most systems exhibit periods in the range of approximately 0.2–0.6 days (\citealt{2003CoSka..33...38P, 2006Ap.....49..358D, 2021ApJS..254...10L, 2024RAA....24a5002P}[3,4,5,6]). A prominent observational feature in many contact binaries is the O’Connell effect, in which the two maxima outside of eclipse in the light curve have unequal brightness (\citealt{1951PRCO....2...85O, 1992ApJ...385..621A}[7,8]). This asymmetry is commonly attributed to surface inhomogeneities such as starspots, enhanced magnetic activity, or mass-transfer streams impacting one of the components.

Mass and energy exchange play a central role in the dynamics and long-term evolution of contact systems. Depending on the direction and rate of mass transfer, the binary may experience angular momentum loss, thermal relaxation oscillations, or transitions between contact, semi-detached, and near-contact phases. Despite decades of investigation, many aspects of their structural configuration, mass transfer behavior, angular momentum evolution, and eventual merger pathways remain open questions (\citealt{2016ApJ...817..133Z}[9]).

Recent large-scale surveys such as Gaia and TESS have significantly increased the number of known contact binaries (\citealt{2021ApJS..254...10L}[5]), yet many systems still lack precise determinations of their fundamental parameters. Accurate measurements of the mass ratio, fillout factor, temperature difference, and fillout factor are critical for distinguishing between stable thermal-contact configurations and systems that are undergoing mass ratio reversal or evolving toward merger (\citealt{2008MNRAS.390.1577G, 2018PASJ...70...90K}[10,11]). Therefore, detailed investigation of individual W UMa systems remains essential for refining empirical correlations, improving population statistics, and testing evolutionary models that predict whether a given system will remain in contact, transition into a semi-detached state, or eventually coalesce into a rapidly rotating single star.

In this work, we perform a detailed light curve analysis of the W UMa-type contact system EM Tuc in order to obtain its physical and orbital parameters and to better constrain its evolutionary configuration. EM Tuc was first reported by \cite{1968BamVe...7...76A}[12] as a variable object. The target is located in the southern hemisphere at $\mathrm{RA.}: 0.506924^{\circ}$ and $\mathrm{DEC.}: -66.88863^{\circ}$ according to Gaia DR3 (\citealt{2023A&A...674A...1G}[13]). EM Tuc has been classified as a contact binary in several major variable star catalogs, including the Variable Star Index (VSX\footnote{\url{https://vsx.aavso.org/}}), the All-Sky Automated Survey for Supernovae (ASAS-SN, \citealt{2018MNRAS.477.3145J}[14]), and Zwicky Transient Facility (ZTF, \citealt{2023A&A...675A.195S}[15]), based on consistent photometric evidence indicating an overcontact configuration. The orbital period listed in the VSX database is $P=0.3265745~\mathrm{d}$, which places EM Tuc among short-period contact binaries. The system reaches a reported maximum brightness of approximately $V \approx 12.05~\mathrm{mag}$, as reported in the VSX database.

This work continues the observational and analytical approach of the BSN\footnote{\url{https://bsnp.info/}} project, focusing on photometric analysis of eclipsing binaries. In this study, we carry out a detailed investigation of the orbital period variations, perform a photometric light-curve analysis, and determine the absolute parameters of the target contact binary EM Tuc.

\vspace{0.4cm}
\section{Observations and data reduction}
Photometric observations of EM Tuc were conducted on 20 September 2023, during which 1053 images were obtained in a single night. The observations were carried out with the 2.15-meter Jorge Sahade (JS) telescope at the Complejo Astronómico El Leoncito (CASLEO) Observatory, San Juan, Argentina (coordinates: $31^\circ48'$ S, $69^\circ18'$ W; altitude 2552 m). A VersArray 2048B cryogenic CCD camera (Roper Scientific, Princeton Instruments) was used, along with standard $BVR_cI_c$ filters. All frames were binned $5 \times 5$ pixels. The exposure times were 25 s for $B$, 12 s for $V$, 12 s for $R_c$, and 7 s for $I_c$ filters.

Standard CCD reductions, including bias subtraction and flat-field correction, were performed on all images. Aperture photometry was carried out using the APPHOT package in IRAF (\citealt{1986SPIE..627..733T}[16]). To correct for airmass effects, a Python routine based on the Astropy package (\citealt{2013A&A...558A..33A}[17]) and following the method of \cite{1962aste.book.....H}[18] was applied. The fluxes were then normalized using AstroImageJ (\citealt{2017AJ....153...77C}[19]).

Time-series photometric data for this study were acquired by the Transiting Exoplanet Survey Satellite (TESS), which is equipped with four wide-field cameras to continuously monitor sections of the sky for exoplanet detection. EM Tuc (TIC 412063998) was observed in sectors 27, 28, and 68, with exposure times of 600 s, 600 s, and 200 s, respectively, and a monitoring duration of approximately 27.4 days per sector. Photometric measurements in the broad “TESS:T” band (600–1000 nm) were retrieved from the Mikulski Archive for Space Telescopes (MAST\footnote{\url{https://mast.stsci.edu/portal/Mashup/Clients/Mast/Portal.html}}). Light-curve processing and data reduction were carried out using the Lightkurve package, applying detrending methods consistent with the Science Processing Operations Center (SPOC) pipeline.

\vspace{0.4cm}
\section{Orbital Period Variations}
A detailed investigation of EM Tuc required a comprehensive collection of eclipse timing measurements, combining archival photometric survey data and our observations. Orbital period variations of this system were examined using these measurements. Published times of minima were collected from various databases and the literature to complement the dataset. Eclipse timings were also directly extracted from the TESS time series, providing consistent and extended coverage where ground-based observations were limited.

This study's multi-filter photometric observations provided additional eclipse timings, enhancing the temporal coverage and strengthening the reliability of the period variation study alongside the data collected from archival surveys. Consistency among all timing measurements was ensured by converting Heliocentric Julian Dates ($HJD$) into Barycentric Julian Dates in Barycentric Dynamical Time ($BJD_{TDB}$; \citealt{2010PASP..122..935E}[20]). An online calculator\footnote{\url{https://astroutils.astronomy.osu.edu/time/hjd2bjd.html}} was employed to perform the conversion. Eclipse timings are provided in Tables \ref{minima} and \ref{tessmin}.

The light elements from the VSX database were adopted as the reference ephemeris for computing the epoch and O–C values (Equation \ref{eq:ref-ephm}).

\begin{equation}
\label{eq:ref-ephm}
BJD_{TDB}(Min.I)=2454290.63344+0.3265745 \times E.
\end{equation}
where $E$ is the number of cycles. 

Using the gathered eclipse timing measurements and the reference ephemeris, a new linear ephemeris was derived for the system, providing an updated basis for computing epochs and O–C values (Equation \ref{eq:new-ephm}).

\begin{equation}
\label{eq:new-ephm}
BJD_{TDB}(Min.I)=2454290.63359(37)+0.32657325(5)\times E.
\end{equation}

\begin{figure*}
\centering
\includegraphics[width=0.7\linewidth]{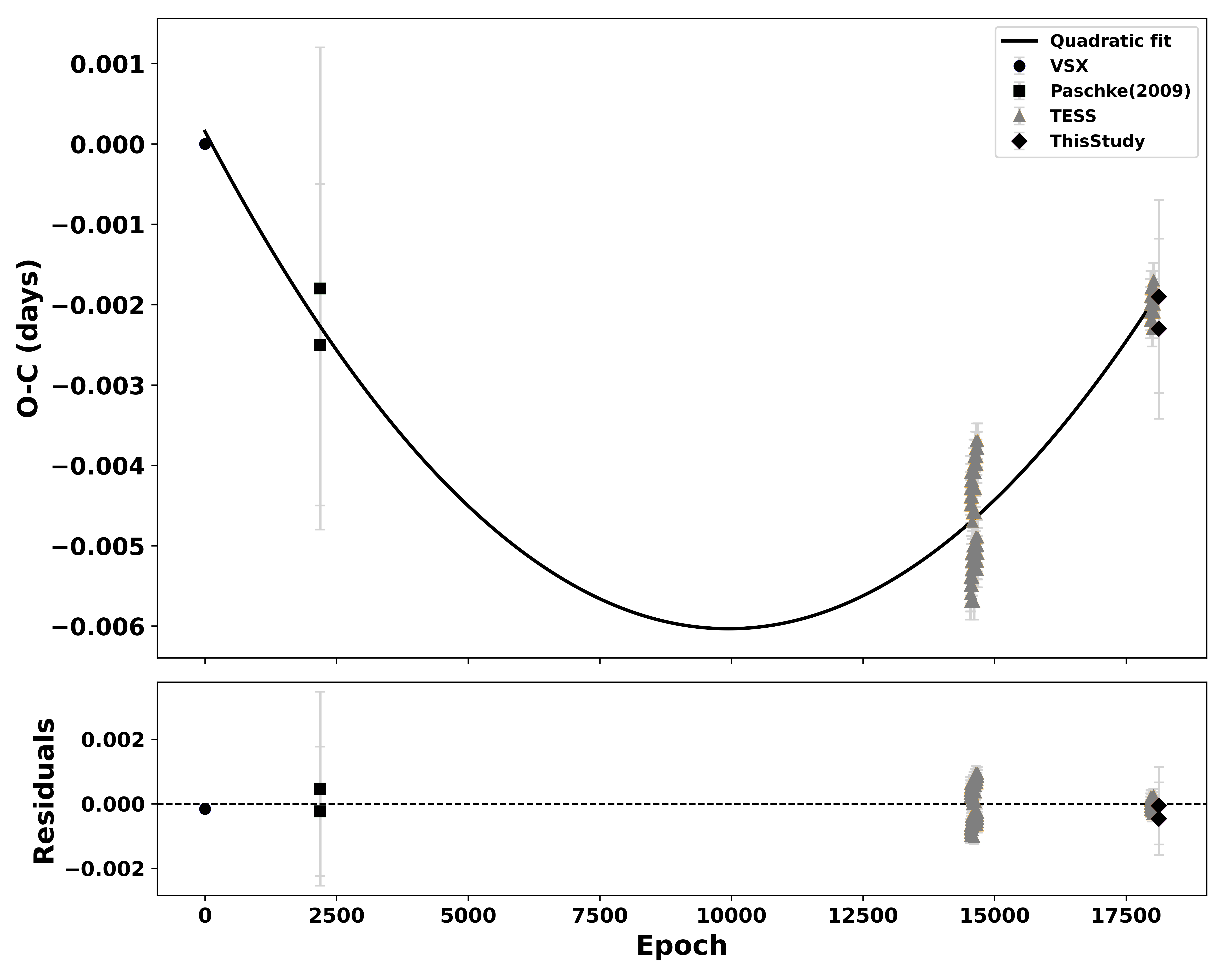}
\caption{The O-C diagram of the EM Tuc contact binary.}
\label{fig:oc}
\end{figure*}

The O–C diagram of EM Tuc is shown in Figure \ref{fig:oc}, together with the corresponding fit residuals. The diagram exhibits an upward parabolic trend, revealing a continuous increase in the orbital period of the system. The derived period change rate is \( dP/dt = (1.401 \pm 0.0421)\times10^{-7}\,\mathrm{d~yr^{-1}} \). Since the hotter component is the less massive star and the cooler component is the more massive one, this positive period derivative indicates that mass is being transferred from the hotter to the cooler component.

\begin{table*}
 \caption{Times of light minimum of ground-based observations.}
 \centering
 \begin{center}
 \begin{tabular}{c c c c c}
\hline
 Min.($BJD_{TDB}$) & Error & Epoch & O-C & Reference\\
\hline
2454290.63344	&		&	0	&	0	&	VSX	\\
2455004.52275	&	0.00200	&	2186	&	-0.0025	&	\cite{2009OEJV..116....1P}	\\
2455004.68675	&	0.00300	&	2186.5	&	-0.0018	&	\cite{2009OEJV..116....1P}	\\
2460208.65139	&	0.00120	&	18121.5	&	-0.0019	& This study\\
2460208.81419	&	0.00112	&	18122	&	-0.0023	& This study\\
\hline
\end{tabular}
\end{center}
\label{minima}
\end{table*}

\vspace{0.4cm}
\section{Parameter Estimations}
\subsection{Light Curve Solution}
\begin{figure}
\centering
\includegraphics[width=0.5\linewidth]{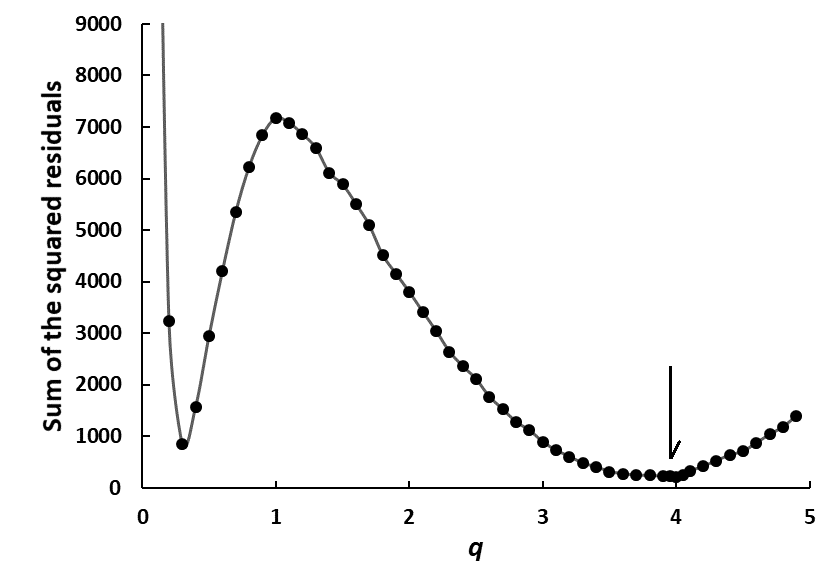}
\caption{The sum of squared residuals as a function of mass ratio. The black arrow indicates the estimated mass ratio in the $q$-search.}
\label{q}
\end{figure}

Light curve modeling of the EM Tuc system was performed with PHOEBE v2.4.9 in combination with an MCMC approach (\citealt{2016ApJS..227...29P, 2020ApJS..250...34C}[22,23]). Based on the overall morphology of the observed light variation and the short orbital period, the configuration of a contact binary was adopted for the solution. The observational times were converted to orbital phase using the updated ephemeris determined in this work (Section 3). The gravity darkening coefficients and bolometric albedos were fixed to $g_1=g_2=0.32$ (\citealt{1967ZA.....65...89L}[24]) and $A_1=A_2=0.5$ (\citealt{1969AcA....19..245R}[25]). Limb-darkening was considered as a free parameter in the process, and stellar atmospheres were represented using the  \cite{2004A&A...419..725C}[26] study. The adopted initial effective temperature of the hotter component follows the value reported in Gaia DR3 (\citealt{2025MNRAS.537.3160P}[27]). The effective temperature of the cooler star was estimated based on the relative depths of the primary and secondary eclipses in the observed light curve. The mass ratio ($q$) is a critical parameter in modeling contact binaries, so a $q$-search was conducted in PHOEBE (\citealt{2005ApSS.296..221T}[28]). A wide interval of mass ratios, spanning from 0.1 to 20, was first examined for each system. After identifying the region that provided a closer agreement with the observations, the search range was gradually reduced (Figure \ref{q}).

Asymmetries between the two light-curve maxima are frequently seen in contact binaries, and such behavior is also evident in our target. This effect is generally interpreted as the result of starspots associated with magnetic activity, commonly referred to as the O'Connell effect (\citealt{1951PRCO....2...85O}[7]).

The light curves of the system were modeled without including a third-light term ($l_3$). No signs of contamination from neighboring sources were identified, and there is currently no strong indication of a third-light contribution.

After determining the most plausible values for main model parameters, they were refined using the PHOEBE's built-in optimization routine. The photometric analysis was further validated using the BSN application (\citealt{2025Galax..13...74P}[29]). The initial version of this application, which runs on Microsoft Windows operating system, offers high processing speed and makes the determination of starspot coordinates considerably easier for the user. Then, the output parameter estimates and their uncertainties were obtained using the Markov Chain Monte Carlo (MCMC) approach implemented with the emcee Python package (\citealt{2013PASP..125..306F}[30]). In this analysis, 96 walkers were initialized around the best-fit solution from the previous optimization step. Each walker was evolved for 1000 iterations, with the first 200 iterations discarded as burn-in to ensure convergence. The resulting posterior distributions were used to determine the median values as the parameter estimates and the 16th and 84th percentiles as their corresponding uncertainties.

\begin{figure*}
\centering
\includegraphics[width=1.0\linewidth]{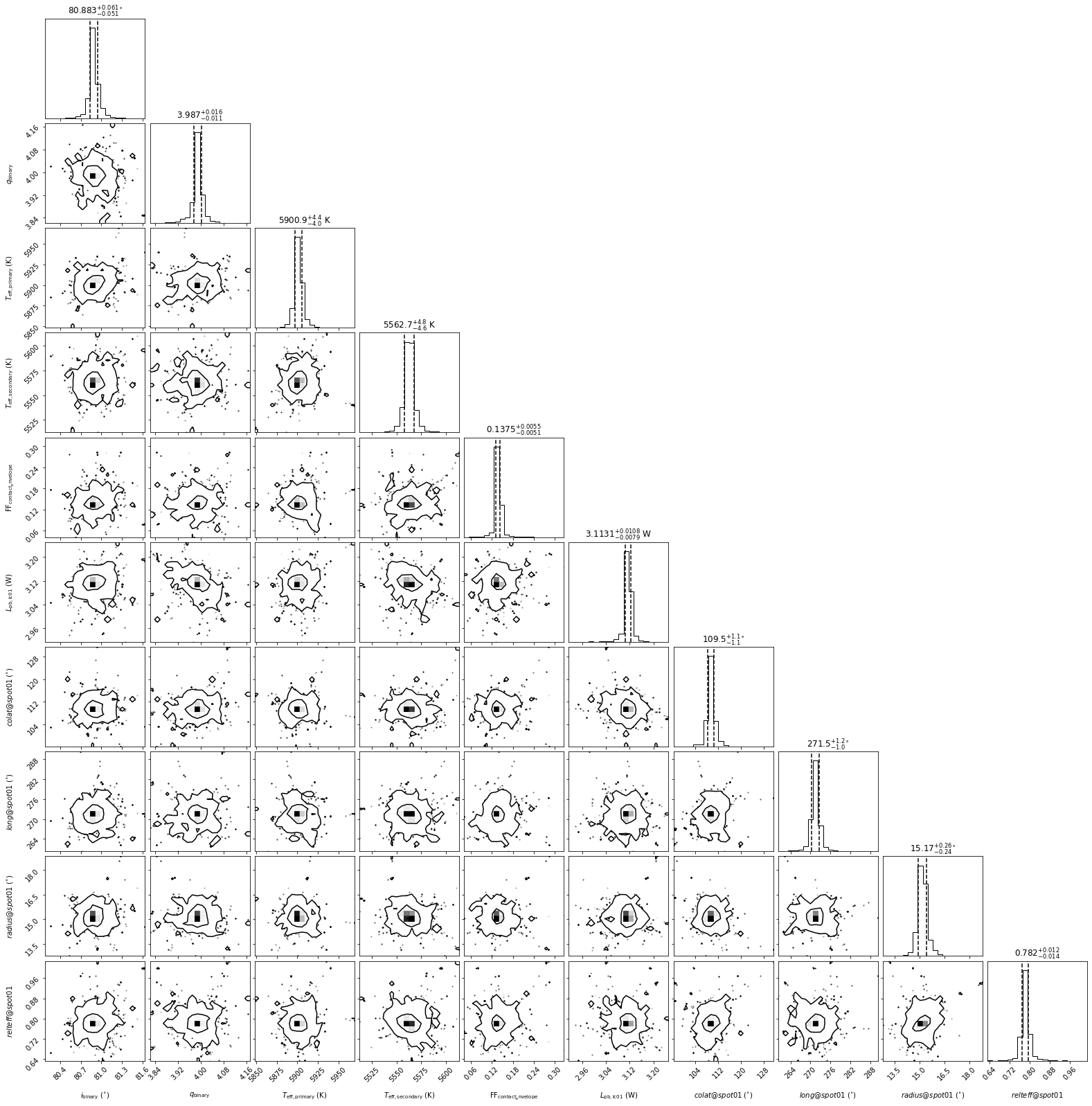}
\caption{The corner plots of EM Tuc determined by MCMC modeling. The parameters used in the MCMC procedure, presented in the corner plot, include: $i$, $q$, $T_1$, $T_2$, $f$, $l_1$, and four starspot parameters, respectively.}
\label{corner}
\end{figure*}

\begin{figure*}
\centering
\includegraphics[width=1\linewidth]{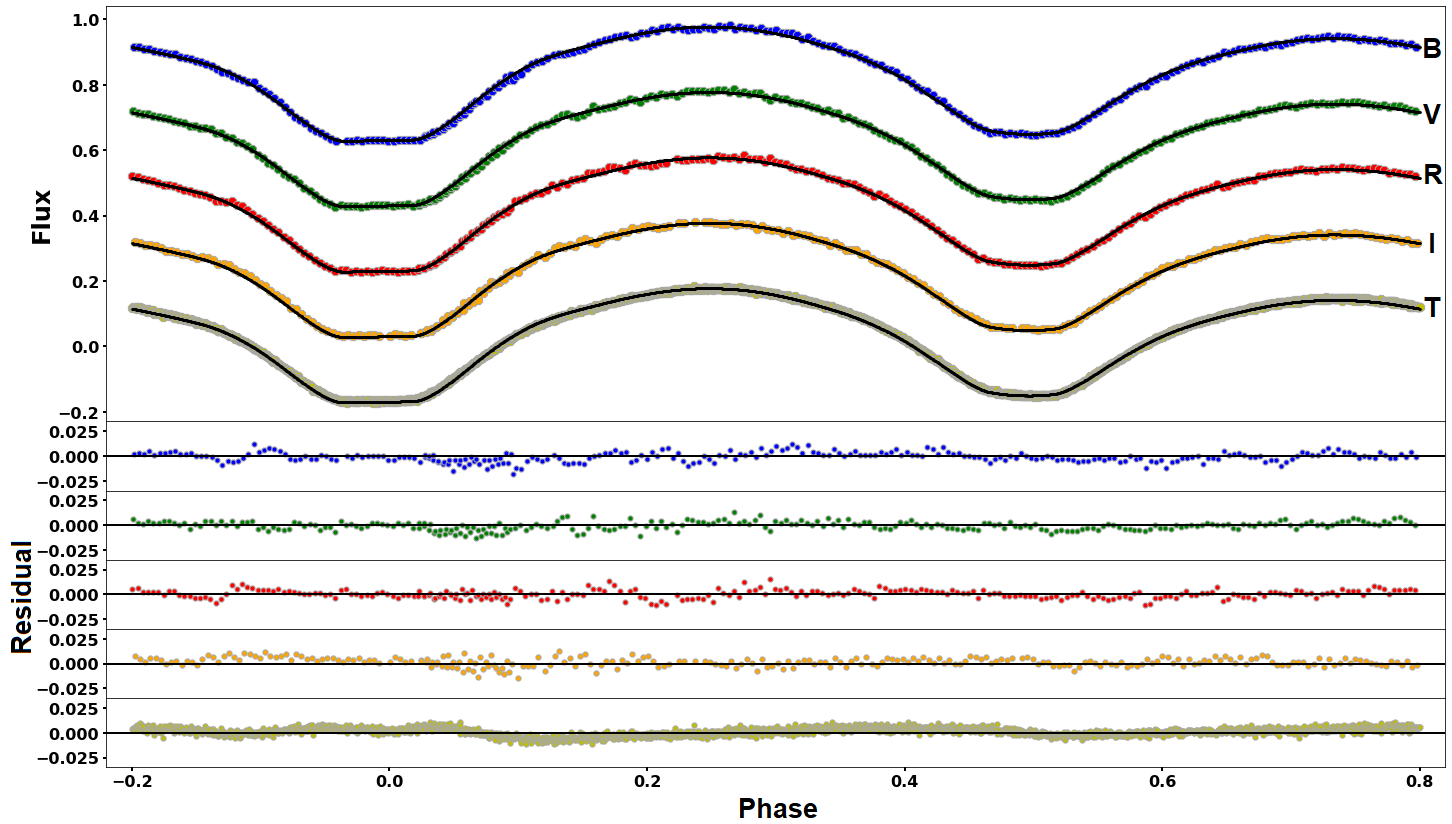}
\caption{Ground- and space-based photometric light curves together with synthetic model.}
\label{lc}
\end{figure*}

\begin{figure*}
\centering
\includegraphics[width=1\linewidth]{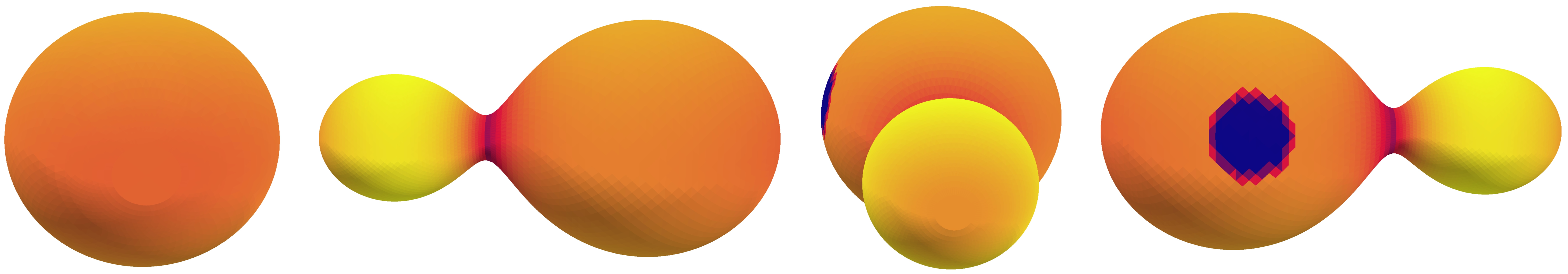}
\caption{3D view of the target system stars.}
\label{3D}
\end{figure*}

Figure \ref{corner} presents the corner plots for the EM Tuc system, illustrating the distributions and correlations of parameters obtained from the MCMC analysis. The final results of the light curve analysis, including starspot characteristics (longitude, colatitude, radius, $T_{\mathrm{spot}}/T_{\mathrm{star}}$), are summarized in Table \ref{analysis}. The observed and synthetic light curves in different filters are displayed in Figure \ref{lc}. Three-dimensional representations of the binary systems, constructed from the final model parameters, are presented in Figure \ref{3D} at orbital phases 0, 0.25, 0.5, and 0.75, respectively.

\begin{table*}
\renewcommand\arraystretch{1.5}
 \caption{Results of light curve solution.}
 \centering
 \begin{center}
 \begin{tabular}{c c c c}
\hline
Parameter & Result & Parameter & Result\\
\hline
$T_h$ (K) 	&	 $5901^{+4}_{-4}$	&	$r_{(mean)_h}$ 	&	 $0.275\pm0.023$\\
$T_c$ (K) 	&	 $5563^{+5}_{-5}$	&	$r_{(mean)_c}$ 	&	 $0.510\pm0.020$\\
$q$ 	&	 $3.987^{+0.016}_{-0.011}$	&	$Col^\circ_{\mathrm{spot}}$ 	&	 110(1)\\
$i^{\circ}$ 	&	 $80.88^{+0.06}_{-0.05}$	&	$Long^\circ_{\mathrm{spot}}$  	&	 272(2)\\
$f$ 	&	 $0.138^{+0.006}_{-0.005}$	&	$Radius^\circ_{\mathrm{spot}}$ 	&	 15(1)\\
$\Omega_h=\Omega_c$ 	&	 $7.808\pm0.315$	&	$T_{\mathrm{spot}}/T_{\mathrm{star}}$ 	&	 0.78(1)\\
$l_h/l_{tot}$ 	&	 $0.262^{+0.001}_{-0.001}$	&	Spot Component 	&	 Cooler star\\
$l_c/l_{tot}$ 	&	 $0.738\pm0.005$	&	MaxI-MaxII (Flux)	& 0.036 \\
\hline
\end{tabular}
\end{center}
\label{analysis}
\end{table*}

\vspace{0.4cm}
\subsection{Absolute Parameters}
The absolute parameters of EM Tuc were derived by using the Gaia DR3 parallax with the photometric solution. Our approach follows the methodology of \cite{2022MNRAS.510.5315P}[31], where the physical properties of both components can be determined once several observational and model-dependent quantities are known. These include the orbital period ($P$), the mean fractional radii ($r_{mean1,2}$), the effective temperatures ($T_{1,2}$), the bolometric corrections ($BC_{1,2}$), the fractional light contributions ($l_{1,2}/l_{tot}$), the interstellar extinction ($A_v$), the system’s maximum brightness ($V_{max}$), and the distance ($d$) from Gaia DR3. Based on these quantities, we derived the absolute visual magnitudes of the system and its individual components, followed by their bolometric magnitudes, luminosities, radii, and masses. The orbital separation was calculated as the average of the two separately determined values, $a_1$ and $a_2$. This method is valid only when $a_1$ and $a_2$ closely agree, ensuring the internal consistency of the solution.

The maximum brightness of EM Tuc was measured as $V_{max}=12.17\pm0.10$ from our photometric light curve. The interstellar extinction was determined as $A_V=0.060\pm0.001$ following the three-dimensional dust maps from \cite{2019ApJ...887...93G}[32], and the system distance was taken as $d = 335(7)$ pc from Gaia DR3. Bolometric corrections of $BC_{1,2}$ were assigned according to the calibration of \cite{1996ApJ...469..355F}[33].

The luminosities and surface gravities were computed using the following commonly adopted relations:
\begin{equation}
M_{\rm bol} - M_{\rm bol,\odot} = -2.5 \log\left(\frac{L}{L_\odot}\right),
\end{equation}
\begin{equation}
g = G_\odot \frac{M}{R^2},
\end{equation}
where solar bolometric magnitude of $M_{\text{bol}\odot} = 4.73$ mag (\citealt{2010AJ....140.1158T}[34]), and $G_\odot$ is the solar gravitational constant.

The resulting physical parameters derived using this parallax-based method are summarized in Table \ref{absolute}.

\begin{table}
\caption{Estimation of absolute parameters.}
\centering
\begin{center}
\footnotesize
\begin{tabular}{c c c}
\hline
Parameter & Hotter Star & Colder Star\\
\hline
$M(M_\odot)$ &  $0.24\pm0.05$ & $ 0.97\pm0.19$ \\
$R(R_\odot)$ & $0.58\pm0.04$ & $1.09\pm0.08$ \\
$L(L_\odot)$ & $0.37\pm0.05$  & $1.03\pm0.14$ \\
$M_{bol}$(mag.) & $5.83\pm0.14$  & $4.71\pm0.14$\\
$log(g)$(cgs) & $4.30\pm0.14$ & $4.35\pm0.15$ \\
$a(R_\odot)$ & \multicolumn{2}{c}{$2.13\pm0.14$} \\
\hline
\end{tabular}
\end{center}
\label{absolute}
\end{table}

\vspace{0.4cm}
\section{Discussion and conclusion}
We carried out ground-based multiband photometric observations of the southern eclipsing binary EM Tuc using the $B$, $V$, $R_c$, and $I_c$ filters. Combining the ground-based data with TESS observations allowed us to perform an in-depth light curve analysis and to study the orbital period variations of the system.

Based on the analysis of the O-C diagram and the quadratic fit, the mass transfer rate in the EM Tuc system has been estimated. In this system, mass is transferred from the less massive component to the more massive one, and this process can be described by the Equation \ref{eq:mass_transfer}:
\begin{equation} \label{eq:mass_transfer}
\dot{M}_1 = - \frac{(dP/P)}{3 \left( \frac{1}{M_1} - \frac{1}{M_2} \right)}.
\end{equation}

The resulting mass transfer rate is \(-4.618 \pm 1.537 \times 10^{-8}~M_\odot\,\mathrm{yr}^{-1}\), indicating that mass is being transferred from the less massive component to the more massive one, which highlights that the system is currently in an active phase of mass exchange.

This mass transfer rate, which lies within the order of \(10^{-8}~M_\odot\,\mathrm{yr}^{-1}\), has significant implications for the physical and evolutionary properties of EM Tuc. The increase in mass of the more massive star and the corresponding decrease of the less massive star lead to changes in the mass ratio and orbital dynamics, potentially affecting the degree of contact and the structural configuration of the components. Furthermore, mass transfer directly influences the stellar internal structure and surface energy distribution, as hydrostatic equilibrium and surface temperature respond to the changing masses, which may also impact the luminosities and radii of both stars. Overall, this process plays a crucial role in the future evolution of the system and is consistent with the characteristics of short-period contact binaries with low mass ratios.

We modeled the light curve using the PHOEBE Python package, and the final parameter values along with their uncertainties were determined through the MCMC approach. The temperature difference between the two stars is 338 K. Based on these temperatures and the spectral classifications provided by \cite{2018MNRAS.479.5491E}[35], the hotter component corresponds to a G2 spectral type, while the cooler one is classified as G8. The light-curve modeling of EM Tuc also required including a cool starspot on the surface of the cooler component to achieve an acceptable fit.

Contact binary systems can be divided into three groups depending on their fillout factor: deep systems with $f \geq 50\%$, medium systems with $25\% \leq f < 50\%$, and shallow systems with $f < 25\%$ (\citealt{2022AJ....164..202L}[36]). According to the light curve modeling, EM Tuc belongs to the shallow class.

The absolute parameters of EM Tuc were estimated using the Gaia DR3 parallax. In this method, the paths of the primary and secondary stars are treated independently, giving semi-major axes $a_1(R_\odot)$ and $a_2(R_\odot)$ for each component. The resulting values are in good agreement, with a small difference of $\Delta a = |a_2 - a_1| = 0.009$, supporting the reliability of the approach. This $\Delta a$ further confirms the validity of the light curve solution and the adopted model parameters (\citealt{2024NewA..11002227P, 2025MNRAS.538.1427P}[37,38]).

\begin{figure*}
\centering
\includegraphics[width=0.96\linewidth]{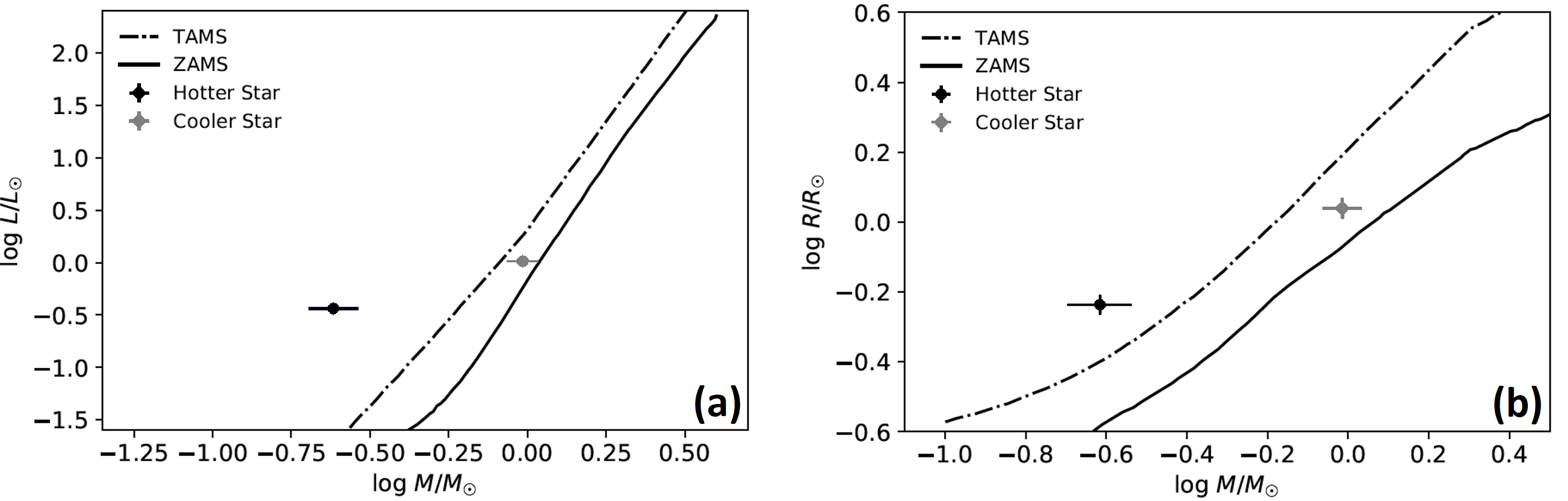}
\caption{$M$-$L$ and $M$-$R$ diagrams of the contact system EM Tuc, showing the evolutionary stages of the components with respect to ZAMS and TAMS.}
\label{relationships}
\end{figure*}

The evolutionary state of EM Tuc was examined through logarithmic Mass-Radius ($M$-$R$) and Mass-Luminosity ($M$-$L$) diagrams, based on the absolute parameters derived for the system (Figures \ref{relationships}a,b). In these diagrams, the locations of both components are compared with the Zero-Age Main Sequence (ZAMS) and Terminal-Age Main Sequence (TAMS) reference tracks from \cite{2000A&AS..141..371G}[39], allowing us to assess their evolutionary progression along the main sequence. As shown in Figures \ref{relationships}a,b, the less massive component is positioned above the TAMS line, whereas the more massive star lies closer to the ZAMS. This indicates that the two components occupy different evolutionary stages along the main sequence.

In this work, EM Tuc is identified as a W UMa-type contact eclipsing binary through light curve modeling using the derived mass ratio, fillout factor, and orbital inclination, which clearly confirm its contact configuration. According to the classification framework for A- and W-type contact binaries proposed by \cite{1970VA.....12..217B}[40], and based on the derived component masses and temperatures (\citealt{2024RAA....24e5001P}[41]), EM Tuc falls into the W-subtype category. In W-type systems, the hotter component is the less massive star, which is consistent with the physical properties obtained for EM Tuc.

\vspace{0.4cm}
\section*{Data Availability}
Ground-based observational data used in this study are available from the corresponding author upon reasonable request.

\vspace{0.4cm}
\section{Acknowledgments}
We used data from the European Space Agency mission Gaia\footnote{\url{http://www.cosmos.esa.int/gaia}}. This work includes data from the TESS mission observations. The NASA Explorer Program provides funding for the TESS mission.

\vspace{1.2cm}
\section*{Appendix}
The appendix table presents the primary and secondary times of minima for the EM Tuc contact binary system, derived from TESS observations. The average uncertainty in the times of minima is 0.00022.

\renewcommand{\thetable}{A\arabic{table}}
\setcounter{table}{0}
\begin{table*}
\renewcommand\arraystretch{0.72}
\caption{Times of Minima extracted from the TESS data.}
\centering
\begin{center}
\begin{tabular}{c c c c c c c c c}
\hline
Min.($BJD_{TDB}$) & Epoch & O-C & Min.($BJD_{TDB}$) & Epoch & O-C & Min.($BJD_{TDB}$) & Epoch & O-C\\
\hline
2459036.40860	&	14532	&	-0.0055	&	2459047.51251	&	14566	&	-0.0051	&	2459059.76011	&	14603.5	&	-0.0040	\\
2459036.57286	&	14532.5	&	-0.0045	&	2459047.67658	&	14566.5	&	-0.0043	&	2459059.92239	&	14604	&	-0.0051	\\
2459036.73525	&	14533	&	-0.0054	&	2459047.83901	&	14567	&	-0.0052	&	2459060.08671	&	14604.5	&	-0.0040	\\
2459036.89952	&	14533.5	&	-0.0044	&	2459048.00328	&	14567.5	&	-0.0042	&	2459060.24885	&	14605	&	-0.0052	\\
2459037.06181	&	14534	&	-0.0054	&	2459049.30916	&	14571.5	&	-0.0046	&	2459060.41321	&	14605.5	&	-0.0041	\\
2459037.22617	&	14534.5	&	-0.0043	&	2459049.47168	&	14572	&	-0.0054	&	2459060.57537	&	14606	&	-0.0052	\\
2459037.38834	&	14535	&	-0.0055	&	2459049.63572	&	14572.5	&	-0.0046	&	2459062.04587	&	14610.5	&	-0.0043	\\
2459037.55267	&	14535.5	&	-0.0044	&	2459049.79826	&	14573	&	-0.0054	&	2459062.20780	&	14611	&	-0.0057	\\
2459037.71496	&	14536	&	-0.0054	&	2459049.96240	&	14573.5	&	-0.0045	&	2459062.53478	&	14612	&	-0.0053	\\
2459037.87931	&	14536.5	&	-0.0044	&	2459050.12493	&	14574	&	-0.0053	&	2459062.69918	&	14612.5	&	-0.0041	\\
2459038.04144	&	14537	&	-0.0055	&	2459050.28880	&	14574.5	&	-0.0047	&	2459062.86136	&	14613	&	-0.0053	\\
2459038.20609	&	14537.5	&	-0.0041	&	2459050.45136	&	14575	&	-0.0054	&	2459063.02602	&	14613.5	&	-0.0039	\\
2459038.36799	&	14538	&	-0.0055	&	2459050.61539	&	14575.5	&	-0.0047	&	2459063.18796	&	14614	&	-0.0052	\\
2459038.53267	&	14538.5	&	-0.0041	&	2459050.77805	&	14576	&	-0.0053	&	2459063.35245	&	14614.5	&	-0.0040	\\
2459038.69450	&	14539	&	-0.0056	&	2459050.94200	&	14576.5	&	-0.0046	&	2459063.51450	&	14615	&	-0.0053	\\
2459038.85924	&	14539.5	&	-0.0041	&	2459051.10457	&	14577	&	-0.0054	&	2459063.67908	&	14615.5	&	-0.0040	\\
2459039.02111	&	14540	&	-0.0056	&	2459051.26854	&	14577.5	&	-0.0047	&	2459063.84110	&	14616	&	-0.0052	\\
2459039.18575	&	14540.5	&	-0.0042	&	2459051.43106	&	14578	&	-0.0054	&	2459064.00568	&	14616.5	&	-0.0039	\\
2459039.34784	&	14541	&	-0.0054	&	2459051.59520	&	14578.5	&	-0.0046	&	2459064.16760	&	14617	&	-0.0053	\\
2459039.51224	&	14541.5	&	-0.0043	&	2459051.75770	&	14579	&	-0.0054	&	2459064.33220	&	14617.5	&	-0.0040	\\
2459039.67411	&	14542	&	-0.0057	&	2459051.92190	&	14579.5	&	-0.0045	&	2459064.49419	&	14618	&	-0.0053	\\
2459039.83886	&	14542.5	&	-0.0042	&	2459052.08421	&	14580	&	-0.0054	&	2459064.65882	&	14618.5	&	-0.0040	\\
2459040.00095	&	14543	&	-0.0054	&	2459052.24852	&	14580.5	&	-0.0044	&	2459064.82091	&	14619	&	-0.0051	\\
2459040.16546	&	14543.5	&	-0.0042	&	2459052.41087	&	14581	&	-0.0054	&	2459064.98538	&	14619.5	&	-0.0040	\\
2459040.32747	&	14544	&	-0.0055	&	2459052.57509	&	14581.5	&	-0.0044	&	2459065.14754	&	14620	&	-0.0051	\\
2459040.49211	&	14544.5	&	-0.0041	&	2459052.73747	&	14582	&	-0.0053	&	2459065.31197	&	14620.5	&	-0.0040	\\
2459040.65407	&	14545	&	-0.0055	&	2459052.90171	&	14582.5	&	-0.0044	&	2459065.47402	&	14621	&	-0.0052	\\
2459040.81860	&	14545.5	&	-0.0042	&	2459053.06389	&	14583	&	-0.0055	&	2459065.63850	&	14621.5	&	-0.0040	\\
2459040.98056	&	14546	&	-0.0056	&	2459053.22822	&	14583.5	&	-0.0044	&	2459065.80078	&	14622	&	-0.0050	\\
2459041.14517	&	14546.5	&	-0.0042	&	2459053.39056	&	14584	&	-0.0054	&	2459065.96519	&	14622.5	&	-0.0039	\\
2459041.30712	&	14547	&	-0.0056	&	2459053.55487	&	14584.5	&	-0.0044	&	2459066.12728	&	14623	&	-0.0051	\\
2459041.47185	&	14547.5	&	-0.0041	&	2459053.71713	&	14585	&	-0.0054	&	2459066.29164	&	14623.5	&	-0.0040	\\
2459041.63370	&	14548	&	-0.0056	&	2459053.88153	&	14585.5	&	-0.0043	&	2459066.45380	&	14624	&	-0.0051	\\
2459041.79835	&	14548.5	&	-0.0042	&	2459054.04366	&	14586	&	-0.0054	&	2459066.61828	&	14624.5	&	-0.0039	\\
2459041.96048	&	14549	&	-0.0054	&	2459054.20809	&	14586.5	&	-0.0043	&	2459066.78043	&	14625	&	-0.0051	\\
2459042.12480	&	14549.5	&	-0.0043	&	2459054.37031	&	14587	&	-0.0054	&	2459066.94488	&	14625.5	&	-0.0039	\\
2459042.28687	&	14550	&	-0.0055	&	2459054.53470	&	14587.5	&	-0.0043	&	2459067.10688	&	14626	&	-0.0052	\\
2459042.45148	&	14550.5	&	-0.0042	&	2459054.69684	&	14588	&	-0.0054	&	2459067.27148	&	14626.5	&	-0.0039	\\
2459042.61344	&	14551	&	-0.0056	&	2459054.86136	&	14588.5	&	-0.0042	&	2459067.43359	&	14627	&	-0.0051	\\
2459042.77801	&	14551.5	&	-0.0043	&	2459055.02351	&	14589	&	-0.0053	&	2459067.59804	&	14627.5	&	-0.0039	\\
2459042.94005	&	14552	&	-0.0055	&	2459055.18790	&	14589.5	&	-0.0042	&	2459067.76011	&	14628	&	-0.0051	\\
2459043.10448	&	14552.5	&	-0.0044	&	2459055.34997	&	14590	&	-0.0054	&	2459067.92469	&	14628.5	&	-0.0038	\\
2459043.26664	&	14553	&	-0.0055	&	2459055.51443	&	14590.5	&	-0.0043	&	2459068.08664	&	14629	&	-0.0052	\\
2459043.43106	&	14553.5	&	-0.0044	&	2459055.67664	&	14591	&	-0.0053	&	2459068.25120	&	14629.5	&	-0.0039	\\
2459043.59327	&	14554	&	-0.0054	&	2459055.84122	&	14591.5	&	-0.0040	&	2459068.41321	&	14630	&	-0.0052	\\
2459043.75770	&	14554.5	&	-0.0043	&	2459056.00332	&	14592	&	-0.0052	&	2459068.57769	&	14630.5	&	-0.0040	\\
2459043.91983	&	14555	&	-0.0055	&	2459056.16781	&	14592.5	&	-0.0040	&	2459068.73983	&	14631	&	-0.0051	\\
2459044.08421	&	14555.5	&	-0.0044	&	2459056.33014	&	14593	&	-0.0050	&	2459068.90437	&	14631.5	&	-0.0039	\\
2459044.24666	&	14556	&	-0.0052	&	2459056.49442	&	14593.5	&	-0.0040	&	2459069.06633	&	14632	&	-0.0052	\\
2459044.41070	&	14556.5	&	-0.0045	&	2459056.65671	&	14594	&	-0.0050	&	2459069.23091	&	14632.5	&	-0.0039	\\
2459044.57329	&	14557	&	-0.0051	&	2459056.82095	&	14594.5	&	-0.0040	&	2459069.39293	&	14633	&	-0.0052	\\
2459044.73761	&	14557.5	&	-0.0041	&	2459056.98321	&	14595	&	-0.0051	&	2459069.55737	&	14633.5	&	-0.0040	\\
2459044.89976	&	14558	&	-0.0053	&	2459057.14750	&	14595.5	&	-0.0041	&	2459069.71949	&	14634	&	-0.0052	\\
2459045.06406	&	14558.5	&	-0.0042	&	2459057.30988	&	14596	&	-0.0050	&	2459069.88389	&	14634.5	&	-0.0041	\\
2459045.22631	&	14559	&	-0.0053	&	2459057.47402	&	14596.5	&	-0.0041	&	2459070.04602	&	14635	&	-0.0052	\\
2459045.39069	&	14559.5	&	-0.0042	&	2459057.63631	&	14597	&	-0.0051	&	2459070.21051	&	14635.5	&	-0.0040	\\
2459045.55294	&	14560	&	-0.0052	&	2459057.80058	&	14597.5	&	-0.0041	&	2459070.37263	&	14636	&	-0.0052	\\
2459045.71712	&	14560.5	&	-0.0043	&	2459057.96293	&	14598	&	-0.0051	&	2459070.53699	&	14636.5	&	-0.0041	\\
2459045.87950	&	14561	&	-0.0052	&	2459058.12723	&	14598.5	&	-0.0041	&	2459070.69928	&	14637	&	-0.0051	\\
2459046.04382	&	14561.5	&	-0.0042	&	2459058.28952	&	14599	&	-0.0050	&	2459070.86359	&	14637.5	&	-0.0041	\\
2459046.20610	&	14562	&	-0.0052	&	2459058.45387	&	14599.5	&	-0.0040	&	2459071.02595	&	14638	&	-0.0050	\\
2459046.37013	&	14562.5	&	-0.0045	&	2459058.61605	&	14600	&	-0.0051	&	2459071.19016	&	14638.5	&	-0.0041	\\
2459046.53265	&	14563	&	-0.0052	&	2459058.78040	&	14600.5	&	-0.0040	&	2459071.35256	&	14639	&	-0.0050	\\
2459046.69675	&	14563.5	&	-0.0044	&	2459058.94266	&	14601	&	-0.0051	&	2459071.51678	&	14639.5	&	-0.0041	\\
2459046.85927	&	14564	&	-0.0052	&	2459059.10706	&	14601.5	&	-0.0039	&	2459071.67927	&	14640	&	-0.0049	\\
2459047.02324	&	14564.5	&	-0.0045	&	2459059.26918	&	14602	&	-0.0051	&	2459071.84357	&	14640.5	&	-0.0038	\\
2459047.18586	&	14565	&	-0.0052	&	2459059.43346	&	14602.5	&	-0.0041	&	2459072.00581	&	14641	&	-0.0049	\\
2459047.34988	&	14565.5	&	-0.0044	&	2459059.59577	&	14603	&	-0.0051	&	2459072.17008	&	14641.5	&	-0.0039	\\
\hline
\end{tabular}
\end{center}
\label{tessmin}
\end{table*}

\renewcommand{\thetable}{A\arabic{table}}
\setcounter{table}{0}
\begin{table*}
\renewcommand\arraystretch{0.72}
\caption{Continued.}
\centering
\begin{center}
\begin{tabular}{c c c c c c c c c}
\hline
Min.($BJD_{TDB}$) & Epoch & O-C & Min.($BJD_{TDB}$) & Epoch & O-C & Min.($BJD_{TDB}$) & Epoch & O-C\\
\hline
2459072.33233	&	14642	&	-0.0049	&	2459082.78254	&	14674	&	-0.0051	&	2460164.23698	&	17985.5	&	-0.0021	\\
2459072.49659	&	14642.5	&	-0.0040	&	2459082.94703	&	14674.5	&	-0.0039	&	2460164.56363	&	17986.5	&	-0.0021	\\
2459072.65899	&	14643	&	-0.0049	&	2459083.10902	&	14675	&	-0.0052	&	2460164.89022	&	17987.5	&	-0.0020	\\
2459072.82315	&	14643.5	&	-0.0040	&	2459083.27370	&	14675.5	&	-0.0038	&	2460165.21678	&	17988.5	&	-0.0021	\\
2459072.98548	&	14644	&	-0.0049	&	2459083.43550	&	14676	&	-0.0053	&	2460165.54334	&	17989.5	&	-0.0021	\\
2459073.14997	&	14644.5	&	-0.0037	&	2459083.60030	&	14676.5	&	-0.0038	&	2460165.87001	&	17990.5	&	-0.0020	\\
2459073.31185	&	14645	&	-0.0051	&	2459083.76217	&	14677	&	-0.0052	&	2460166.19653	&	17991.5	&	-0.0020	\\
2459073.47601	&	14645.5	&	-0.0043	&	2459083.92693	&	14677.5	&	-0.0037	&	2460166.52305	&	17992.5	&	-0.0021	\\
2459073.63896	&	14646	&	-0.0046	&	2459084.08861	&	14678	&	-0.0053	&	2460166.84959	&	17993.5	&	-0.0021	\\
2459073.80296	&	14646.5	&	-0.0039	&	2459084.25346	&	14678.5	&	-0.0038	&	2460167.17620	&	17994.5	&	-0.0021	\\
2459075.27123	&	14651	&	-0.0052	&	2459084.41554	&	14679	&	-0.0050	&	2460167.50256	&	17995.5	&	-0.0023	\\
2459075.43586	&	14651.5	&	-0.0039	&	2459084.58016	&	14679.5	&	-0.0037	&	2460167.82922	&	17996.5	&	-0.0022	\\
2459075.59787	&	14652	&	-0.0051	&	2459084.74191	&	14680	&	-0.0052	&	2460168.15588	&	17997.5	&	-0.0021	\\
2459075.76236	&	14652.5	&	-0.0039	&	2459084.90668	&	14680.5	&	-0.0037	&	2460168.80909	&	17999.5	&	-0.0021	\\
2459076.25083	&	14654	&	-0.0053	&	2459085.06871	&	14681	&	-0.0050	&	2460169.13562	&	18000.5	&	-0.0021	\\
2459076.41569	&	14654.5	&	-0.0038	&	2459085.23325	&	14681.5	&	-0.0037	&	2460169.46212	&	18001.5	&	-0.0022	\\
2459076.57781	&	14655	&	-0.0049	&	2459085.39521	&	14682	&	-0.0050	&	2460169.78880	&	18002.5	&	-0.0021	\\
2459076.74218	&	14655.5	&	-0.0038	&	2459085.55980	&	14682.5	&	-0.0037	&	2460170.11539	&	18003.5	&	-0.0021	\\
2459076.90435	&	14656	&	-0.0050	&	2459085.72188	&	14683	&	-0.0049	&	2460170.44204	&	18004.5	&	-0.0020	\\
2459077.06879	&	14656.5	&	-0.0038	&	2459086.04835	&	14684	&	-0.0051	&	2460170.76852	&	18005.5	&	-0.0021	\\
2459077.23066	&	14657	&	-0.0052	&	2459086.21292	&	14684.5	&	-0.0038	&	2460171.09507	&	18006.5	&	-0.0021	\\
2459077.39542	&	14657.5	&	-0.0038	&	2459086.37489	&	14685	&	-0.0051	&	2460171.42165	&	18007.5	&	-0.0021	\\
2459077.55731	&	14658	&	-0.0052	&	2459086.53958	&	14685.5	&	-0.0037	&	2460171.74831	&	18008.5	&	-0.0020	\\
2459077.72191	&	14658.5	&	-0.0038	&	2459086.70146	&	14686	&	-0.0051	&	2460172.07501	&	18009.5	&	-0.0019	\\
2459077.88391	&	14659	&	-0.0051	&	2459086.86604	&	14686.5	&	-0.0038	&	2460172.40160	&	18010.5	&	-0.0019	\\
2459078.04854	&	14659.5	&	-0.0038	&	2459087.02804	&	14687	&	-0.0051	&	2460172.72809	&	18011.5	&	-0.0020	\\
2459078.21044	&	14660	&	-0.0052	&	2460154.76623	&	17956.5	&	-0.0022	&	2460173.05475	&	18012.5	&	-0.0019	\\
2459078.37507	&	14660.5	&	-0.0038	&	2460155.09292	&	17957.5	&	-0.0021	&	2460173.38129	&	18013.5	&	-0.0019	\\
2459078.53691	&	14661	&	-0.0053	&	2460155.41962	&	17958.5	&	-0.0020	&	2460173.70797	&	18014.5	&	-0.0018	\\
2459078.70162	&	14661.5	&	-0.0039	&	2460155.74607	&	17959.5	&	-0.0021	&	2460174.03453	&	18015.5	&	-0.0018	\\
2459078.86355	&	14662	&	-0.0052	&	2460156.07282	&	17960.5	&	-0.0019	&	2460174.36100	&	18016.5	&	-0.0019	\\
2459079.02821	&	14662.5	&	-0.0038	&	2460156.39935	&	17961.5	&	-0.0020	&	2460174.68762	&	18017.5	&	-0.0019	\\
2459079.19019	&	14663	&	-0.0051	&	2460156.72597	&	17962.5	&	-0.0019	&	2460175.34099	&	18019.5	&	-0.0017	\\
2459079.35483	&	14663.5	&	-0.0038	&	2460157.05248	&	17963.5	&	-0.0020	&	2460175.66745	&	18020.5	&	-0.0018	\\
2459079.51669	&	14664	&	-0.0052	&	2460157.37922	&	17964.5	&	-0.0018	&	2460175.99406	&	18021.5	&	-0.0017	\\
2459079.68132	&	14664.5	&	-0.0039	&	2460157.70571	&	17965.5	&	-0.0019	&	2460176.32062	&	18022.5	&	-0.0017	\\
2459079.84329	&	14665	&	-0.0052	&	2460158.03225	&	17966.5	&	-0.0019	&	2460176.64715	&	18023.5	&	-0.0018	\\
2459080.00801	&	14665.5	&	-0.0038	&	2460158.35881	&	17967.5	&	-0.0020	&	2460176.97370	&	18024.5	&	-0.0018	\\
2459080.16984	&	14666	&	-0.0052	&	2460158.68542	&	17968.5	&	-0.0019	&	2460177.30032	&	18025.5	&	-0.0018	\\
2459080.33444	&	14666.5	&	-0.0039	&	2460159.01201	&	17969.5	&	-0.0019	&	2460177.62688	&	18026.5	&	-0.0018	\\
2459080.49646	&	14667	&	-0.0052	&	2460159.33858	&	17970.5	&	-0.0019	&	2460177.95336	&	18027.5	&	-0.0019	\\
2459080.66098	&	14667.5	&	-0.0039	&	2460159.66523	&	17971.5	&	-0.0018	&	2460178.27996	&	18028.5	&	-0.0019	\\
2459080.82301	&	14668	&	-0.0052	&	2460159.99179	&	17972.5	&	-0.0019	&	2460178.60642	&	18029.5	&	-0.0020	\\
2459080.98752	&	14668.5	&	-0.0040	&	2460160.31831	&	17973.5	&	-0.0019	&	2460178.93320	&	18030.5	&	-0.0018	\\
2459081.14959	&	14669	&	-0.0052	&	2460160.64496	&	17974.5	&	-0.0018	&	2460179.25970	&	18031.5	&	-0.0018	\\
2459081.31414	&	14669.5	&	-0.0039	&	2460160.97149	&	17975.5	&	-0.0019	&	2460179.58619	&	18032.5	&	-0.0019	\\
2459081.47621	&	14670	&	-0.0051	&	2460161.29813	&	17976.5	&	-0.0018	&	2460179.91276	&	18033.5	&	-0.0019	\\
2459081.64060	&	14670.5	&	-0.0040	&	2460161.62455	&	17977.5	&	-0.0020	&	2460180.23924	&	18034.5	&	-0.0020	\\
2459081.80275	&	14671	&	-0.0052	&	2460162.27760	&	17979.5	&	-0.0021	&	2460180.56589	&	18035.5	&	-0.0019	\\
2459081.96725	&	14671.5	&	-0.0040	&	2460162.60435	&	17980.5	&	-0.0019	&	2460180.89254	&	18036.5	&	-0.0019	\\
2459082.12920	&	14672	&	-0.0053	&	2460162.93085	&	17981.5	&	-0.0020	&	2460181.54550	&	18038.5	&	-0.0021	\\
2459082.29382	&	14672.5	&	-0.0040	&	2460163.25741	&	17982.5	&	-0.0020	&	2460181.87211	&	18039.5	&	-0.0020	\\
2459082.45588	&	14673	&	-0.0052	&	2460163.58398	&	17983.5	&	-0.0020	&		&		&		\\
2459082.62044	&	14673.5	&	-0.0039	&	2460163.91036	&	17984.5	&	-0.0022	&		&		&		\\
\hline
\end{tabular}
\end{center}
\end{table*}


\end{document}